\documentclass{aa}
\usepackage{txfonts}
\usepackage{graphicx}
\usepackage{multicol}
\usepackage{multirow}
\usepackage{natbib}
\bibpunct{(}{)}{;}{a}{}{,}

\newcommand\blfootnote[1]{%
  \begingroup
  \renewcommand\thefootnote{}\footnote{#1}%
  \addtocounter{footnote}{-1}%
  \endgroup
}

\begin{document}

   \title{Chromospheric activity in bright contact binary stars}
   
   \authorrunning{T. Mitnyan, T. Szalai, A. B\'odi et al.}
   \titlerunning{Chromospheric activity in contact binaries}

   \author{T. Mitnyan\inst{1,2}, T. Szalai\inst{1,3},  A. B\'odi\inst{3,4}, L. Kriskovics\inst{3}, K. Vida\inst{3}, B. Cseh\inst{3}, O. Hanyecz\inst{3}, A. Ordasi\inst{3}, A. P\'al\inst{3}, J. Vink\'o\inst{1,3}
          }

   \institute{Department of Optics and Quantum Electronics, University of Szeged, H-6720 Szeged, D\'om t\'er 9, Hungary
\and 
Baja Astronomical Observatory of University of Szeged, H-6500 Baja, Szegedi \'ut, Kt. 766, Hungary\\
             \email{mtibor@titan.physx.u-szeged.hu}
\and
             Konkoly Observatory, Research Centre for Astronomy and Earth Sciences,  H-1121 Budapest, Konkoly Thege Mikl\'os \'ut 15-17, Hungary
             \and
             MTA CSFK Lend\"ulet Near-Field Cosmology Research Group, Konkoly Thege Mikl\'os \'ut 15-17, H-1121 Budapest, Hungary
             }
             
   \date{Received 29 November 2019; accepted 27 January 2020}

% \abstract{}{}{}{}{} 
% 5 {} token are mandatory
 
  \abstract
   {Studying chromospheric activity of contact binaries is an important way of revealing the magnetic activity processes of these systems. An efficient but somewhat neglected method for that is to follow the changes of the H$\alpha$ line profiles via optical spectroscopy.}
   {Our goal was to perform a comprehensive preliminary analysis based on the optical spectral signs of chromospheric activity on the largest sample of contact binaries to date.}
   {We collected optical echelle spectra on 12 bright contact binaries at 17 nights. We derived new radial velocity curves from our observations. For quantifying the apparent chromospheric activity levels of the systems, we subtracted self-constructed synthetic spectra from the observed ones and measured the equivalent widths of the residual H$\alpha$-profiles at each observed epoch. Our well-sampled data set allowed us to study the short-term variations of chromospheric activity levels as well as to search for correlations between them and some basic physical parameters of the systems.}
   {Fitting the radial velocity curves, we re-determined the mass ratios and systemic velocities of all observed objects. We found that chromospheric activity levels of the studied systems show various changes during the orbital revolution: we see either flat, or one-peaked, or two-peaked distributions of equivalent width vs. the orbital phase. The first case means that the activity level is probably constant, while the latter two cases suggest the presence of one or two active longitudes at the stellar surfaces. Our correlation diagrams show that mean chromospheric activity levels may be in connection with orbital periods, B$-$V color indices, inverse Rossby numbers, and temperature differences of the components. At the same time, no clear trend is visible with respect to mass ratios, inclinations and fill-out factors of the systems. A- and W-type contact binaries in our sample show similar distributions at each of the studied correlation diagrams.}
   {}


   \maketitle

\blfootnote{Based on data collected with 2-m RCC telescope at Rozhen National Astronomical Observatory, Bulgaria.}

\section{Introduction}

Stellar activity is a key phenomenon in studying close binary stars: it may (strongly) affect several observable of these systems and may also influence their long-term evolution. It can induce angular momentum redistribution that may play a role in the formation and evolution of these systems and may lead to cyclic variations in their orbital periods \citep[e.g.][]{applegate92,lanza04}.

Contact binary systems consist of two low-mass main-sequence stars (mostly F, G, or K spectral type) that are in a physical contact with each other through the inner (L1) Lagrangian point. The orbital period of these binaries is usually less than a day; the shape of the components is strongly distorted, which causes continuous variation in the observed light curves (LCs). According to the most accepted model, the components are embedded in a common convective envelope that ensures mass and energy transport between them \citep{lucy68}.

Considering their LCs, these objects can be separated into two main types: A-type and W-type \citep{binnendijk70}. In A-type systems, the more massive (from now primary) component has higher surface brightness than the less massive (from now secondary) one, while in W-type systems, the primary has lower surface brightness than the secondary. There are two more additional categories: B-type \citep{lucy79} and H-type systems \citep{csizmadia04}. In B-type systems, the components are not in thermal contact and have a temperature difference higher than 1000\,K, while H-type systems have mass ratios higher than 0.72.

Many contact binaries show activity signals. One of these signals is the asymmetry in the LC maxima (O'Connell effect), which is most likely caused by starspots on the surface of either or both components \citep[e.g.][]{mullan75}. Emission excess observed in certain ultraviolet (UV), optical, and/or infrared (IR) absorption spectral lines (e.g. Mg II, H$\alpha$, Ca II) probably hint at ongoing chromospheric activity \citep[e.g.][]{rucinski85,barden85,montes00} of the stars, while detection of X-ray emission gives us insight into coronal activity \citep[e.g.][]{cruddace84,vilhu84}. Taken all together, observing different forms of stellar magnetic activity is essential to get a full picture of the physical processes behind it.

In this paper, we focus on the chromospheric activity of contact binaries detected as excess emission in the optical H$\alpha$ line. This method is somewhat neglected in the literature because it is complicated to disentangle the chromospheric and photospheric effects on the rotationally broadened line profiles of these systems. \citet{rucinski85} analyzed the Mg II emission in UV spectra of some W UMa binaries in order to extend the relation found between the strength of chromospheric activity and the inverse Rossby number for non-contact stars \citep{noyes84,hartmann84}. He found that chromospheric activity is not a monotonic function of either the orbital (i.e. rotational) period or the B--V color index. At the same time, the inverse Rossby number has a strong correlation with the level of chromospheric activity and roughly follows the relation found for non-contact stars with slower rotation. \citet{barden85} analyzed the optical spectra of some RS CVn and W UMa systems aiming to find similar relations. After subtracting the photospheric parts of H$\alpha$ lines, he defined the emission excess as the degree of chromospheric activity level. In his preliminary study, he also found a correlation between the level of chromospheric activity and inverse Rossby number. In order to get information on the individual activity levels of the binary components, he fitted the sums of two Gaussians on the residual spectra. He showed that, in contact binaries, there is a shutdown in the activity of the secondary components with decreasing rotational velocity. He concluded that the possible reasons for this shutdown can be either the common envelope, or the tidal interaction, or the combination of both. He also emphasized that this correlation should be studied in details on more systems.

Our motivation was to extend the preliminary results of \citet{barden85} via performing a similar analysis on a larger sample of contact binaries. The paper is organized as follows: in Sect. 2., we present all the practical information about data reduction and analysis methods. In Sect. 3., we analyze the results for each individual objects and discuss the nature of the observed chromospheric activities with respect to some fundamental astrophysical parameters. Finally, in Sect. 4., we summarize our work and give the concluding remarks of our study.

\section{Observations and analysis}

\subsection{Spectroscopy}

We collected most of our data between March 2018 and July 2019 (on 17 nights in total) using the (R $\sim$ 20\,000) echelle
spectrograph mounted on the 1m RCC telescope of Konkoly Observatory, Hungary. Additionally, we also collected data in 4 nights of observations using the 2m RCC telescope equipped with the (R $\sim$ 30\,000) ESpeRo spectrograph \citep{bonev17} at the National Astronomical Observatory Rozhen, Bulgaria. The observed objects were chosen according to their apparent brightness and visibility, because the used instruments allowed us to observe only the brightest contact binaries with short exposure times -- in order to avoid smearing of the line profiles -- and reasonable signal-to-noise ratio (SNR). We could not cover the full orbital cycle of every object with measurements because of our limited observing time and of weather conditions. Detailed log of our observations can be found in Table \ref{observations}. Data reduction was performed by standard \texttt{IRAF}\footnote{Image Reduction and Analysis Facility: http://iraf.noao.edu} procedures using \texttt{ccdred} and \texttt{echelle} packages. We regularly took bias, dark, and flat images for the corrections of instrumental effects, and ThAr spectral lamp spectra for wavelength calibration. For the continuum normalization, we applied a two-step process: {\it i)} we constructed the blaze function from the flat images for every echelle order, then the original spectral orders were divided by their estimated blaze function; {\it ii)} we used the built-in method of \texttt{iSpec} \citep{blanco14, blanco19} to fit the remnant deviations from the continuum with low-order splines for each spectral order and, after that, we stitched them together to construct the 1-D spectra. Finally, we applied barycentric correction and telluric line removal on every spectrum applying \texttt{iSpec} routines.

\begin{table*}
\caption{Log of spectroscopic observations.}
\centering
\begin{tabular}{c c c c c c}
\hline \hline
Object (Type) & Dates of observations & Observatory & Number of spectra & Integration time (s) & Average SNR\\ 
\hline
KR Com (A) & 22,23 March 2018 & Konkoly & 35 & 900, 1200 & 67\\
V1073 Cyg (A) & 27,28 August 2018 & Konkoly & 36 & 1200 & 46\\
V2150 Cyg (A) & 29 August 2018 & Konkoly & 21 & 1200 & 48\\
LS Del (W) & 21 August 2018 & NAO Rozhen & 37 & 600 & 54\\
V972 Her (W) & 27,29,30 April 2018 & Konkoly & 28 & 900 & 124\\
SW Lac (W) & 22 August 2018 & NAO Rozhen & 42 & 600 & 52\\
EX Leo (A) & 27 March; 27,28 April 2018 & Konkoly & 27 & 1200 & 69\\
V351 Peg (A) & 27,28 September 2018 & Konkoly & 41 & 1200 & 96\\
V357 Peg (A) & 21,23 August 2018 & NAO Rozhen & 25 & 1200 & 55\\
OU Ser (W) & 13,14 June 2019 & Konkoly & 39 & 720 & 47\\
V781 Tau (W) & 26 December 2018 & Konkoly & 33 & 900 & 47\\
HX UMa (A) & 17 January 2019 & Konkoly & 30 & 900 & 67\\
\hline
\end{tabular}
\label{observations}
\end{table*}

\subsection{Radial velocities}

We applied the cross-correlation method for deriving the radial velocities (RVs) of the components. The cross-correlation functions (CCFs) were computed with \texttt{iSpec} for every spectra using the built-in NARVAL Sun spectrum as a template. The SNR of our spectra decreases significantly at shorter wavelengths, while there is some fringing at higher wavelengths. Thus, we decided to use the wavelength range of 4800 $- $ 6500 \mbox{\AA}, which was free from both of these effects, for further analysis. The typical CCFs show two or three wide and blended peaks, which were fitted by sum of Gaussians in order to determine the positions of their maxima. We eliminated spectra close to the eclipsing orbital phases (where the components are close to each other and their CCF profiles are so heavily blended that they cannot be distinguished). A sample of the CCFs close to the second quadrature ($\phi\sim0.75$) with the corresponding Gaussian fits are plotted for every object in Fig. \ref{ccf_all}. After constructing the RV curves of the objects, we fitted them with \texttt{PHOEBE} \citep{prsa05} assuming circular orbits in order to get constraints on mass ratios ($q$) and gamma velocities ($V_{\gamma}$). We used the formal errors of the Gaussian peaks as the standard deviation of the RV points. The strong correlation between semi-major axis ($a$) and inclination ($i$) does not allow to get information on both parameters independently, hence we chose to use previously determined $i$ values from the literature and fixed them during the fitting process. Thus, only $a$, $q$, and $V_{\gamma}$ were free parameters. We also enabled a phase shift during the fittings in order to check the accuracy of the applied HJD zero points and orbital periods.

\subsection{Spectrum synthesis}

\begin{table*}
\caption{Input parameters from the literature. HJD$_\mathrm{0}$ and $P$ values are from \citet{kreiner04} for all systems, except V972 Her \citep{rucinski02}. Almost every study misses uncertainty values for one or more of the collected parameters, hence we decided to show the errors for $q$ and V$_{\gamma}$ only in order to be able to compare them with our results. References: 1) \citet{selam04}, 2) \citet{zasche10}, 3) \citet{rucinski02}, 4) \citet{tian18}, 5) \citet{pribulla06},  6) \citet{kreiner03}, 7) \citet{lu01}, 8) \citet{deb11}, 9) \citet{lu99}, 10) \citet{selam18}, 11) \citet{gazeas05}, 12) \citet{senavci11}, 13) \citet{zola10}, 14) \citet{albayrak05}, 15) \citet{rucinski01}, 16) \citet{rucinski08}, 17) \citet{rucinski00}, 18) \citet{kallrath06}, 19) \citet{selam05}, 20) \citet{rucinski03}}
\centering
\begin{tabular}{c c c c c c c c c c}
\hline \hline
Object (Type) & $HJD_\mathrm{0}$ & $P$ [d] & $i$ [\degr] & $f$ & $T_\mathrm{eff,1}$ [K] & $T_\mathrm{eff,2}$ [K] & $q$ & V$_{\gamma}$ & Ref.\\
\hline
KR Com (A) & 2452500.3920 & 0.4079676 & 52.14 & 0.70 & 6072 & 5549 & 0.091 (2) & $-$7.86 (38) & 1, 2, 3\\
V1073 Cyg (A) & 2452500.4776 & 0.7858492 & 68.40 & 0.12 & 7300 & 6609 & 0.303 (17) & $-$6.85 (50) & 4, 5\\
V2150 Cyg (A) & 2452500.5220 & 0.5918576 & 43.39 & 0.21 & 8000 & 7920 & 0.802 (6) & $-$12.82 (45) & 6, 7\\
LS Del (W) & 2452500.3431 & 0.3638427 & 45.25 & 0.09 & 6192 & 6250 & 0.375 (10) & $-$25.90 (14) & 8, 9\\
V972 Her (W) & 2451349.1808 & 0.4430940 & 40.07 & 0.01 & 6046 & 6522 & 0.164 (14) & $+$4.55 (70) & 3, 10\\
SW Lac (W) & 2452500.0690 & 0.3207256 & 79.80 & 0.30 & 5515 & 5800 & 0.781 (7) & $-$10.34 (65) & 11, 12\\
EX Leo (A) & 2452500.2160 & 0.4086068 & 60.80 & 0.65 & 6340 & 6110 & 0.199 (36) & $-$11.05 (1.10) & 7, 13\\
V351 Peg (A) & 2452500.4948 & 0.5932974 & 63.00 & 0.21 & 7559 & 7580 & 0.360 (6) & $-$8.08 (89) & 14, 15\\
V357 Peg (A) & 2452500.3021 & 0.5784510 & 73.23 & 0.10 & 7000 & 6438 & 0.401 (4) & $-$10.84 (54) & 8, 16\\
OU Ser (W) & 2452500.0650 & 0.2967682 & 50.47 & 0.68 & 5940 & 5759 & 0.173 (17) & $-$64.08 (41) & 8, 17\\
V781 Tau (W) & 2452500.0739 & 0.3449097 & 65.89 & 0.205 & 5804 & 6000 & 0.405 (11) & $+$25.74 (1.85) & 18\\
HX UMa (A) & 2452500.1027 & 0.3791546 & 48.85 & 0.59 & 6650 & 6601 & 0.291 (9) & $-$19.88 (1.11) & 19, 20\\
\hline
\end{tabular}
\label{input_parameters}
\end{table*}

In order to get quantitative information about the chromospheric activity levels, we compared the observed spectra of each system to synthetic ones. Since, in contact binaries, the temperatures of the components can be assumed to be nearly equal, we used the assumption that the spectra of these binaries can be represented by a single Doppler-shifted and rotationally broadened model atmosphere. We synthesized \texttt{MARCS.GES} model atmospheres \citep{gustafsson08} using solar abundances \citep{grevesse07} with the \texttt{SPECTRUM} radiative transfer code \citep{gray94} in \texttt{iSpec}. The Doppler-shifted model atmospheres were convoluted with theoretical broadening functions (BFs) to construct simple models of each observed spectra; BFs were calculated with the \texttt{WUMA4} program \citep{rucinski73}. Basic parameters of stars used for the calculations and their references are summarized in Table \ref{input_parameters}. We assumed solar log $g$ and metallicity values for the model atmospheres and varied only the effective temperatures to get the best-fit models. We omitted the H$\alpha$ region during the fitting process, hence we could get information about the level of chromospheric emission filling in the core of the H$\alpha$ line. We also applied a careful smoothing with a low (0.5 \mbox{\AA}) FWHM Gaussian, which left the spectral lines intact, but significantly lowered the noise level in our observed spectra before the fitting process. After the fitting, we simply subtracted the models from the observed spectra and measured the equivalent widths (EWs) of the residual H$\alpha$ emission profiles in a 10\mbox{\AA} wide window using \texttt{IRAF}/\texttt{splot}. We applied the direct integration method to calculate the EWs, because we usually could not yield satisfactory fits using either Gaussian or Voigt profiles. For estimating the uncertainties of the determined EW values, we tested the effects of two potential sources: i) not perfect continuum normalization of the observed spectra, and ii) (known) uncertainties of adopted physical parameters of the systems. We found that the previous effect may produce much larger propagated errors, thus, we estimated the uncertainties of EW values by calculating the RMS of the relative differences between the observed and synthetic spectra in the fitted regions.

\subsection{List of targets}

Our targets can be separated into two categories: i) long-known, well-studied systems with published combined (photometric and spectroscopic) analyses (KR Com, V1073 Cyg, LS Del, SW Lac, and V781 Tau); ii) systems that are relatively neglected in the literature (V2150 Cyg, V972 Her, EX Leo, V351 Peg, V357 Peg, OU Ser, and HX UMa).

For most of these objects, signs of chromospheric activity have not been directly detected yet; the two exceptions are SW Lac \citep{rucinski85} and HX UMa \citep{kjurkchieva10}. Nevertheless, some of these systems show other stellar activity signals such as i) night-to-night variations and unequal maxima in the LCs indicating spots on the surface of the components 
(V1073 Cyg -- \citealt{yang00}; V2150 Cyg -- \citealt{yesilyaprak02}; LS Del -- \citealt{demircan91}, \citealt{derman91}; SW Lac -- \citealt{gazeas05}, \citealt{senavci11}; EX Leo -- \citealt{pribulla02b}, \citealt{zola10}; V351 Peg -- \citealt{albayrak05}; V357 Peg -- \citealt{ekmekci12}; OU Ser -- \citealt{pribulla02}, \citealt{yesilyaprak02}; V781 Tau -- \citealt{cereda88}, \citealt{kallrath06}, \citealt{li16}), ii) long-term modulation in the orbital period, which could be connected to magnetic cycles (KR Com -- \citealt{zasche10}; V1073 Cyg -- \citealt{pribulla06}; V781 Tau -- \citealt{li16}), or, iii) X-ray flux coming from the direction of the system indicating coronal activity (KR Com -- \citealt{kiraga12}; LS Del -- \citealt{stepien01}, \citealt{szczygiel08}, \citealt{kiraga12}; SW Lac -- \citealt{cruddace84}, \citealt{mcgale96}, \citealt{stepien01}, \citealt{xing07}; EX Leo -- \citealt{kiraga12}; OU Ser -- \citealt{kiraga12}; V781 Tau -- \citealt{stepien01}, \citealt{kiraga12}).

In the cases of V1073 Cyg and V781 Tau,  \citet{pribulla06} and \citet{li16} showed that the observed long-term period variations can not be explained with magnetic cycles; instead, the authors proposed that they might be caused by a light-time effect (LITE) of faint tertiary components "hiding" in these systems. Additional components are very common in contact binaries \citep{pribulla06b} and assumed to play a key role in the formation and evolution of these systems \citep{eggleton01}. In our sample, two binaries have directly detected bright third components (KR Com -- \citealt{zasche10}; HX UMa -- \citealt{rucinski03}), while, in another three systems the observed long-term period variations are most likely caused by LITE of an undetected tertiary component (V1073 Cyg, SW Lac, V781 Tau). SW Lac may consist of even more than three stars: \citet{yuan14}, based on a detailed period analysis, showed that there are signals of at least three more components in this system (C, D, E). They note that the distant visual component discovered by \citet{rucinski07} can be most likely matched with component E. After estimating the minimal masses for components C and D, they proposed that the faint spectral contribution reported by \citet{hendry98} belongs most likely to component C or E. They also stated that component D should be visible in the spectra unless it is a compact object or does not exist at all. We note that we did not detect signs for any other components in our spectra beyond the close binary.

For the two objects with measurable third-light contribution in the observed spectra (KR Com, HX UMa), we had to mimic the effect of these additional components in our synthetic spectra. For doing that, model atmospheres of T$_\mathrm{eff}$=6000\,K were convoluted with a Gaussian corresponding to the transmission function of the used spectrograph and were simply added to the binary models giving 30$\%$ \citep{rucinski02,zasche10} and 5$\%$ \citep{rucinski03} of the total luminosity of KR Com and HX UMa, respectively. For KR Com, \citet{zasche10} estimated the effective temperature of the tertiary as T$_\mathrm{eff}$=5900 $\pm$ 200\,K, and we chose T$_\mathrm{eff}$=6000\,K according to that. For HX UMa, we do not have such kind of an estimated temperature from earlier studies, hence, we tried a wide range of T$_\mathrm{eff}$ values during the modeling. We found that it did not affect the spectra significantly, because of the very low fraction of the third light to the total luminosity (5$\%$). Models with the tertiary T$_\mathrm{eff}$ around 6000\,K gave the best solutions, hence, we chose this value for the final solution, but again, it causes only minor differences in the composite spectra. In the spectra of V1073 Cyg and V781 Tau, as in the case of SW Lac, we did not find any spectral signs for the possible additional components, hence we did not take into account any of them during the construction of synthetic spectra.

We also note that in the cases of three objects (SW Lac, V357 Peg, V781 Tau), physical parameters (especially the effective temperatures) adopted from the most recent analyses \citep[][respectively]{senavci11,ekmekci12,li16} did not result in well-fitted models of our observed spectra. In these cases, we decided to adopt physical parameters from earlier studies \citep[][respectively]{gazeas05,deb11,kallrath06}, which allowed us to construct better models.  

\section{Results}

\begin{table*}
\caption{Parameters determined from radial velocity curve modeling and comparison of the derived and published $(M_1+M_2)\,\mathrm{sin}^3i$ values. The references are the same as in Table 2.}.
\centering
\begin{tabular}{c c c c c | c c}
\hline \hline
Object & Phase shift & $a\,\mathrm{sin}i$ [R$_{\odot}$] & $V_{\gamma}$ [km s$^{-1}$] & $q$ & $(M_1+M_2)\,\mathrm{sin}^3i$ & $(M_1+M_2)\,\mathrm{sin}^3i$\\
 &  &  &  &  & This study & Reference value\\
\hline
KR Com & $-$0.0209 (42) & 1.87 (2) & $-$7.54 (65) & 0.072 (4) & 0.527 (17) & 0.517 (8)\\
V1073 Cyg & +0.0276 (62) & 4.60 (6) & $-$5.59 (1.10) & 0.284 (6) & 2.127 (82) & 1.896 (25)\\
V2150 Cyg & $-$0.4677 (71) & 3.22 (8) & $-$8.01 (3.66) & 0.790 (45) & 1.283 (95) & 1.376 (18)\\
LS Del & $-$0.0286 (29) & 1.92 (2) & $-$25.06 (71) & 0.391 (5) & 0.714 (22) & 0.617 (12)\\
V972 Her & +0.0275 (24) & 1.58 (1) & +3.10 (30) & 0.168 (2) & 0.272 (5) & 0.276 (6)\\
SW Lac & $-$0.0242 (40) & 2.59 (2) & $-$5.93 (1.51) & 0.785 (13) & 2.256 (52) & 2.101 (55)\\
EX Leo & $-$0.0064 (33) & 2.42 (2) & $-$7.24 (1.07) & 0.190 (5) & 1.142 (28) & 1.255 (36)\\
V351 Peg & $-$0.4886 (43) & 3.80 (5) & $-$12.23 (1.30) & 0.410 (9) & 2.084 (82) & 2.214 (57)\\
V357 Peg & +0.0016 (51) & 3.93 (6) & $-$17.68 (2.11) & 0.355 (13) & 2.429 (111) & 2.112 (18)\\
OU Ser & +0.0173 (44) & 1.55 (2) & $-$60.86 (1.32) & 0.142 (8) & 0.569 (22) & 0.640 (9)\\
V781 Tau & $-$0.4057 (29) & 2.19 (2) & +31.73 (1.04) & 0.399 (7) & 1.191 (32) & 1.275 (147)\\
HX UMa & +0.0147 (59) & 2.07 (4) & +7.54 (1.67) & 0.274 (11) & 0.834 (48) & 0.775 (30)\\
\hline
\end{tabular}
\label{radseb}
\end{table*}

Our newly derived RV curves and the corresponding PHOEBE model fits are shown in Appendix A (Fig. \ref{rv_all}). The parameters determined from the fitted curves are summarized in Table \ref{radseb}. In the cases of three systems (V2150 Cyg, V351 Peg and V781 Tau), corrections were needed for the calculated orbital phases according to the large phase shift values. For the other objects, the applied HJD reference points and orbital periods seem to be correct. The derived mass ratios and systemic velocities are mostly consistent with the values found in previous studies, if we take into consideration that our uncertainties are only lower limits as the formal errors of the fitted parameters, hence, can be underestimated. In some cases, small differences ($\la\pm$5 km s$^{-1}$) can be seen in the systemic velocities, which are most likely caused by the combined effect of moderate SNR and orbital phase coverage. Larger differences ($\ga\pm$5 km s$^{-1}$) in the systemic velocities compared to published values may indicate the presence of tertiary components in the systems. This is the case for HX UMa, V357 Peg, and V781 Tau, where the first one has a known tertiary, while the latter two have not.

A sample of observed and synthetic spectra of every object is presented in Appendix B (Fig. \ref{models_all}). In most cases, because of the blended profiles, it is difficult to separate the effects of the two components in the H$\alpha$ line emission excess seen on the residual spectra. Because of that reason, we consistently measured the cumulative EW values of the components in every case.

\subsection{Short-term variations in the chromospheric activity}

\begin{figure*}
   \centering
   \includegraphics[width=17cm]{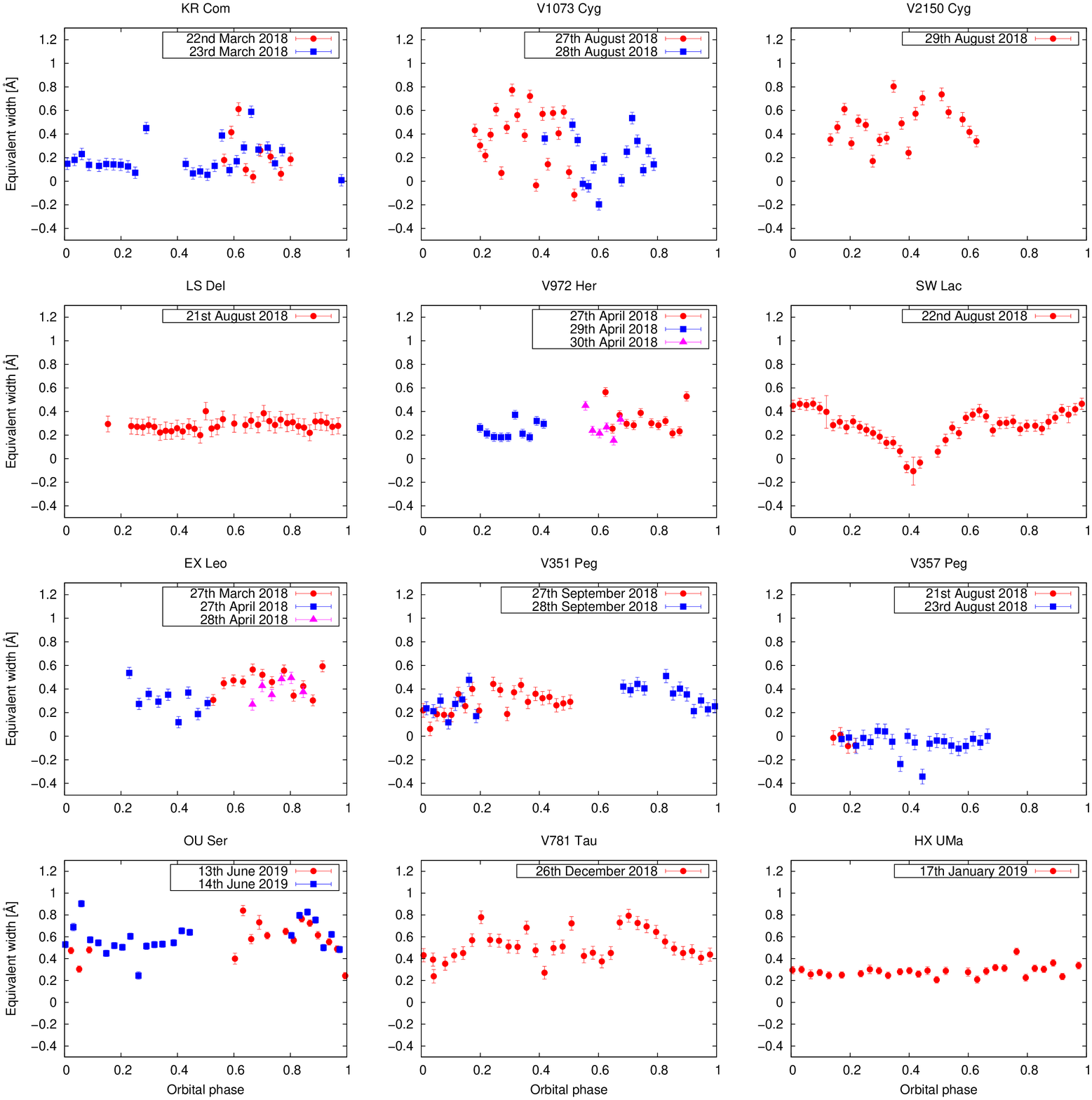}
   \caption{The measured EWs of the residual H$\alpha$ emission line profile of all observed objects versus orbital phase.}
   \label{ew-phase_all}
   \end{figure*}

The measured EWs, representing the apparent chromospheric activity level, as the function of the orbital phases ($\phi$) are plotted in Fig. \ref{ew-phase_all}. In the residual spectra, H$\alpha$ profiles show emission in almost every case, therefore the measured EWs are negative. We decided to use the absolute value of these numbers, because this convention better expresses the chromospheric activity level and its variation. In a few cases, H$\alpha$ profiles show a net absorption in the residual spectra (therefore the measured EWs should be positive, but, because of the convention mentioned above, we handle them as negative numbers). Observations carried out on different nights are marked with different colors and symbols. EW values of the individual systems measured on different nights are mostly consistent with each other. Our spectroscopic phase coverage varies between $\sim50-100\%$. The observed variability of the chromospheric activity level can be categorized into three groups: i) flat distribution -- mostly constant activity level during the orbital revolution; ii) one-peak distribution -- enhanced activity level at a specific orbital phase; iii) two-peak distribution -- enhanced activity levels at two different orbital phases. According to previous studies, enhanced chromospheric activity might be connected to spots on the surfaces of either or both components \citep[e.g.][]{kaszas98,mitnyan18}.

The objects showing a flat distribution of the EW values are LS Del, V972 Her, V357 Peg, and HX UMa. These systems seem to have constant chromospheric activity levels during their orbital revolution. In the case of V972 Her, there seem to be some narrow peaks in the EW-phase diagram; however, these are probably caused by some misdetermined values (e.g. because of unsatisfactory continuum fit). Moreover, the orbital phase coverage is only $\sim$70$\%$, hence the category of this binary is uncertain.

Systems which EW values show one-peak distribution are KR Com, V1073 Cyg, SW Lac, EX Leo and OU Ser. These objects have enhanced chromospheric activity level at a certain orbital phase. This might be explained by the presence of a large spot or a group of spots visible on the surface of either or both components faced us at the given orbital phase. In the cases of KR Com and OU Ser, there is significant scattering in the data around the observed peak, thus, it is doubtful whether those peaks are real or not. We note that V1073 Cyg and EX Leo might have another peak around $\phi$=0.2 and $\phi$=0.9, respectively, which is not covered by our observations.

Objects with a two-peak distribution of the EW values are V2150 Cyg, V351 Peg, and V781 Tau. These systems have enhanced chromospheric activity level at two different orbital phases. This might indicate two active regions on the surface of one of the components \citep[as in the case of e.g. VW Cep, see][]{mitnyan18}. For V351 Peg and V781 Tau, the two peaks are located approximately a half orbital phase from each other. This is consistent with the model of \citet{holzwarth03}, who showed that, in active close binaries, spots are most likely distributed on the opposite sides of the stars because of the presence of strong tidal forces. We note that in the case of V1073 Cyg and EX Leo, if there is also a peak which is uncovered with observations, than it is also located about a half orbital phase from the observed peak. V2150 Cyg does not follow this trend, however, it has the worst orbital coverage ( $\sim$50\%) in our sample, hence, its peak distribution is the most uncertain.

\subsection{Correlation diagrams}

After analyzing the short-term variations of the chromospheric activity levels of the individual systems, we examined the possible connections between the average chromospheric activity levels and some fundamental parameters of the stars. We also  included VW Cephei in our sample, which data were analyzed in a similar way in our previous paper \citep{mitnyan18}. However, in order to be consistent, we re-analyzed that data set exactly in the same way as described in Sect. 2.3. Our final sample (13 stars) is still not large, but it is a significant increment with respect to the 4 stars analyzed by \citet{barden85}. Moreover, \citet{barden85} analyzed only W-type contact binaries and completely avoided A-type systems, while our sample contains almost equal numbers of both types. This means that we can perform a comprehensive analysis concerning this matter, however, the results will be still preliminary and can be biased by some effects such as sample selection or activity cycles. These effects could be ruled out only by significantly increasing the sample size in the future.

   \begin{figure}
   \centering
   \includegraphics[width=9cm]{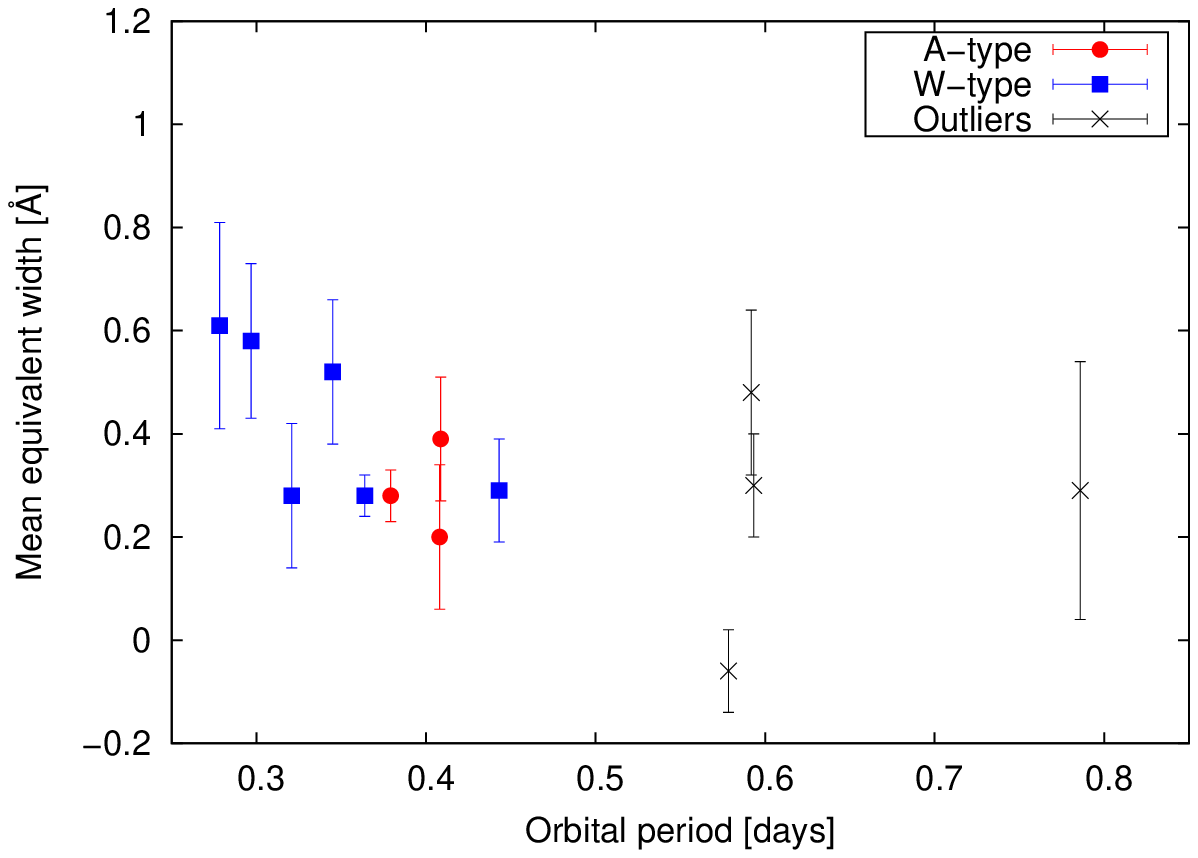}
   \caption{The chromospheric activity level averaged for the whole orbital cycle versus the orbital period of the system.}
   \label{ew-porb}
   \end{figure}

   \begin{figure}
   \centering
   \includegraphics[width=9cm]{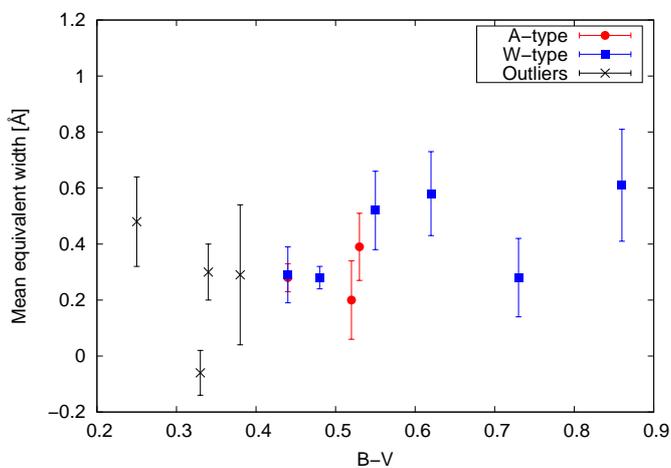}
   \caption{The chromospheric activity level averaged for the whole orbital cycle versus the B--V color index of the system.}
   \label{ew-BV}
   \end{figure}
   
   \begin{figure}
   \centering
   \includegraphics[width=9cm]{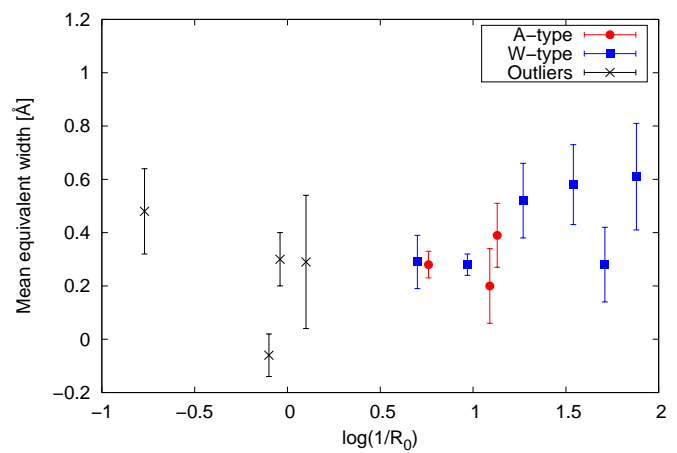}
   \caption{The chromospheric activity level averaged for the whole orbital cycle versus the logarithm of the inverse Rossby-number of the system.}
   \label{ew-loginvR0}
   \end{figure}
   
   \begin{figure}
   \centering
   \includegraphics[width=9cm]{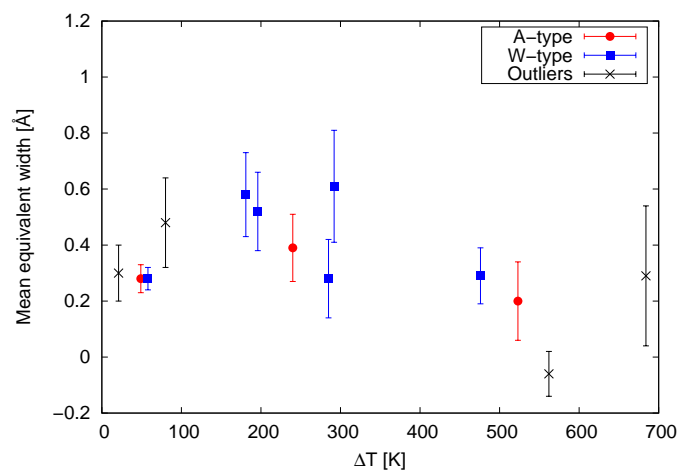}
   \caption{The chromospheric activity level averaged for the whole orbital cycle versus the temperature difference of the components in the system.}
   \label{ew-dT}
   \end{figure}
   
   \begin{figure}
   \centering
   \includegraphics[width=9cm]{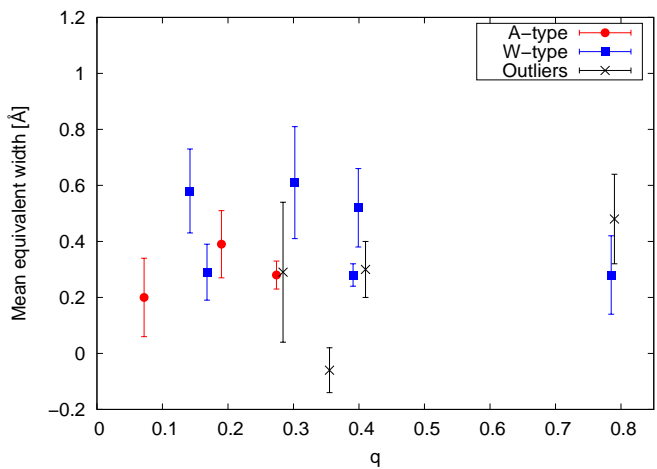}
   \caption{The chromospheric activity level averaged for the whole orbital cycle versus the mass ratio of the system.}
   \label{ew-q}
   \end{figure}

   \begin{figure}
   \centering
   \includegraphics[width=9cm]{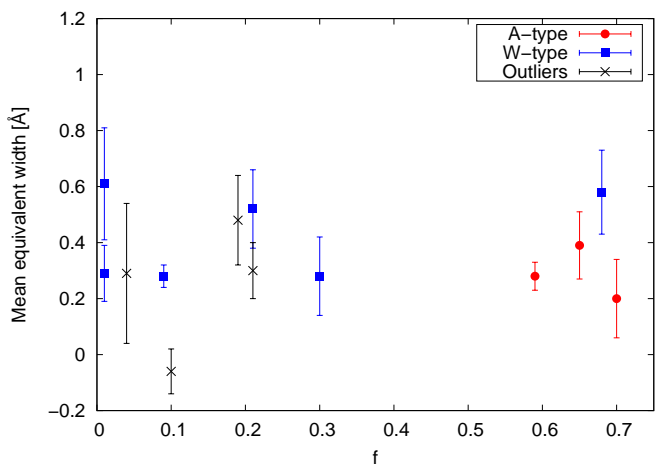}
   \caption{The chromospheric activity level averaged for the whole orbital cycle versus the fill-out factor of the system.}
   \label{ew-f}
   \end{figure}
   
   \begin{figure}
   \centering
   \includegraphics[width=9cm]{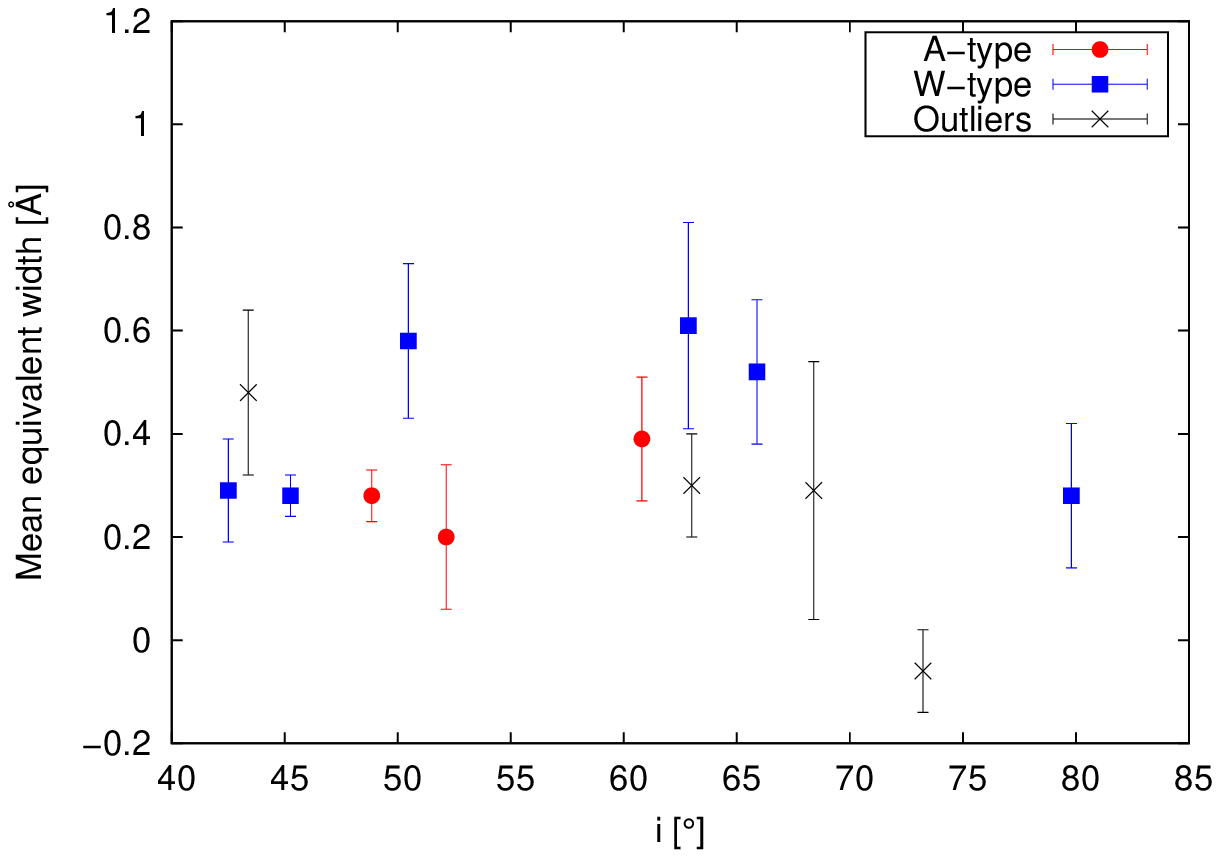}
   \caption{The chromospheric activity level averaged for the whole orbital cycle versus the inclination of the system.}
   \label{ew-i}
   \end{figure}
   
For creating correlation diagrams, we averaged all the measured EW values of the residual H$\alpha$ emission profiles for every single system in our sample. Then we plotted these mean EW values with respect to several fundamental parameters of the systems in order to find possible correlations (Figs. \ref{ew-porb}-\ref{ew-i}.). The errors are given as the standard deviations of the individual EW values from the means. There are 4 objects (V1073 Cyg, V2150 Cyg, V351 Peg and V357 Peg) in our diagrams which do not follow the same trends as the other part of the sample. They have very common properties: they have the longest orbital periods, the smallest B--V color indices and the smallest inverse Rossby numbers. Moreover, they all belong to the A-type contact binaries. Actually, V357 Peg is the only one out of these four systems that shows almost zero chromospheric emission as it is expected. However, the other three objects have significant emission, which is unexpected and does not fit in the trend that the other part of the sample shows. We propose that, for these systems, there may be another source of the H$\alpha$ emission excess instead of the chromospheric activity. These objects are marked with black crosses in Figs. \ref{ew-porb}-\ref{ew-i} as outliers and are left out from the following discussions.

Fig. \ref{ew-porb} shows mean chromospheric activity level (i.e. mean EWs, called hereafter MCAL) versus orbital period for each system. It seems that the system with the shortest orbital period has the most active chromosphere. The activity level drops significantly towards longer orbital periods up to $\sim$0.45 days. Both A- and W-type binaries in our sample follow this relation. Systems with short orbital periods ($\lesssim$0.45 days) are more represented in our correlation diagram than that of with longer periods; thus, we can put stronger constrains only for the previous systems and further observations are needed for studying the latter ones. 

MCAL versus B--V color index is presented in Fig. \ref{ew-BV}. It shows that the system with the highest B--V value is the most chromospherically active and activity level decreases towards smaller B--V values. This trend continues up to B--V$\simeq$0.45. Both A- and W-type systems seem to follow this trend. Objects with B--V$\lesssim$0.65 are better sampled, while more observations are needed in order to better populate the redder part of the correlation diagram.

In Fig. \ref{ew-loginvR0}, MCAL is plotted versus the logarithm of the inverse Rossby number (see Appendix C for the derivation method of the Rossby numbers). It shows a very similar trend to that of B--V correlation diagram (as it is expected since the value of the Rossby number is calibrated by the B--V color index), but it is more clear in this representation. Again, both types of contact binaries follow the same trend.

We also analyzed the possible connection of the activity level and the temperature differences of the binary components. MCAL versus $T_{\mathrm{eff}}$ values in Fig. \ref{ew-dT} show a slight increase up to $\simeq$200K, where they peak and then decrease back to approximately the original level. There seem to be no outstanding differences between A- and W-type systems.

In Fig. \ref{ew-q}, MCAL is plotted against the mass ratio of the systems. It shows a somewhat scattered, but mostly flat distribution without any clear trends. We note that there are no objects in our sample with 0.45 $\lesssim q \lesssim$ 0.75, which also makes it more difficult to identify possible trends. 

Next, we plotted MCAL versus the fill-out factors ($f$) of the systems. Fig. \ref{ew-f} shows a flat distribution similar to the previous diagram (with a smaller scattering). There is also a gap in our data for 0.3 $\lesssim f \lesssim$ 0.6, similarly to the case of mass ratios.

In our last correlation diagram (Fig. \ref{ew-i}), we show MCAL versus the inclinations of the systems. This shows a scattered, but mostly flat distribution similarly to the previous two diagrams. Again, there is no significant difference between the behavior of A- and W-type systems.

\section{Summary}

In this study, we presented the results of new optical spectroscopic observations on 12 contact binary systems. Our motivation was to i) get information about the short-term variations of the chromospheric activity levels and ii) to analyze the possible connections between the mean chromospheric activity level and some fundamental stellar parameters. Based on our own spectral data set, we derived new RV curves applying the cross-correlation technique and modeled these RV curves with \texttt{PHOEBE} to re-determine mass ratios and systemic velocities of the objects. Our newly derived $q$ and $V_{\gamma}$ values are mostly consistent with previous ones found in the literature. 

In order to get constraints on the chromospheric activity level, we constructed synthetic spectra for all observed phases of the observed binaries. After that, we subtracted the model spectra from the observed ones and measured the EW values in a 10\,\AA \,wide window centered on the residual H$\alpha$ line. We used these EW values as a representation of the apparent chromospheric activity strength and analyzed their variations during the orbital cycle. Regarding most of the studied systems (excluding SW Lac and HX UMa), this is the first time that any direct signs of chromospheric activity are detected. We also found different kinds of short-term variations such as continuously changing activity level with one or two peaks, or constant activity level during the whole orbital revolution. More spectra with better SNR and better temporal resolution -- combined with simultaneous photometric observations -- could allow to study both short- and long-term variations of the chromospheric activity more deeply.

For the purpose of searching for correlations between the chromospheric activity level and the fundamental parameters of the observed systems, we averaged all EW values of the individual systems measured in different orbital phases. After re-analyzing in the same way, we also included VW Cep in our sample from our previous paper \citep{mitnyan18}. Based on our correlation diagrams, there is a clear connection between the mean chromospheric activity level of contact binaries and their orbital periods, their B--V color indices and their inverse Rossby numbers. This is consistent with earlier results based on the emission of UV Mg II line  \citep{rucinski85} and also with that of based on the analysis of optical H$\alpha$ lines of a smaller sample of contact binaries \citep{barden85}. We also found a possible trend on the diagram showing the mean chromospheric activity level against the temperature difference of the components. Both A- and W-type systems follow these trends in a similar way. At the same time, we did not find any clear trends in MCAL versus mass ratio, fill-out factor, or the inclination of the systems.

In order to get these conclusions, we had to exclude 4 objects (V1073 Cyg, V2150 Cyg, V351 Peg and V357 Peg) with very similar physical properties from the correlation analysis. We suppose that, in these systems, some other mechanism may be responsible for the observed H$\alpha$ emission excess. A possible explanation could be the mass transfer between the components, which was previously proposed by \citet{beccari14} for some contact binaries with unusually large H$\alpha$ EWs.

While our sample of contact binaries is the largest to date for which chromospheric activity has been studied via optical spectral signs, all of these conclusions are still preliminary and should be considered with caution. The results can be biased e.g. with selection effects, or magnetic activity cycles. A further increment of the sample size is needed for ruling out these effects efficiently and also to better constrain and quantify these relations.

\section*{Acknowledgments}

We would like to thank to our anonymous referee for his/her comments which helped us to further improve the quality of this paper. This project has been supported by the Lendület grants LP2012-31 and LP2018-7/2019 of the Hungarian Academy of Sciences, by the Hungarian National Research, Development and Innovation Office, NKFIH KH-130372 grant and by the \'UNKP-18-3 New National Excellence Program of the Ministry of Human Capacities. TM gratefully acknowledge observing grant support from the Institute of Astronomy and Rozhen National Astronomical Observatory, Bulgarian Academy of Sciences. TM would like to thank to Ventsislav Dimitrov, Grigor Nikolov, Milen Minev, Lyubomir Simeonov, Mitko Churalski and Ilian Iliev for their generous support they gave during the observations in Bulgaria.

\bibliographystyle{aa}
\bibliography{main}

\begin{appendix}

\section{CCFs and radial velocity curves}

   \begin{figure*}
   \centering
   \includegraphics[width=17cm]{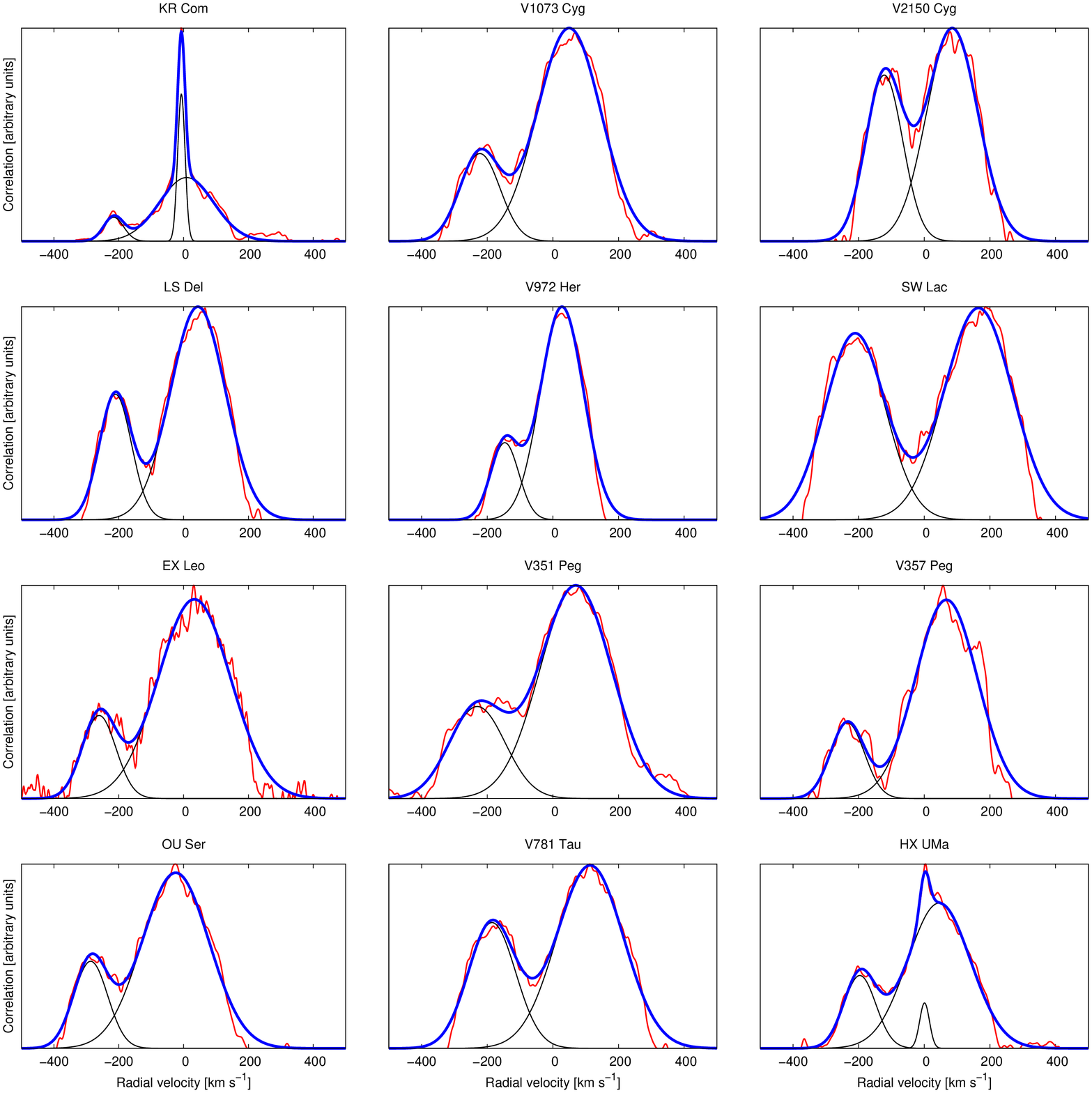}
   \caption{A sample of the derived CCFs (red) for every object close to the second quadrature ($\phi\sim0.75$) with the fitted Gaussians (black) and their sum (blue).}
   \label{ccf_all}
   \end{figure*}

   \begin{figure*}
   \centering
   \includegraphics[width=17cm]{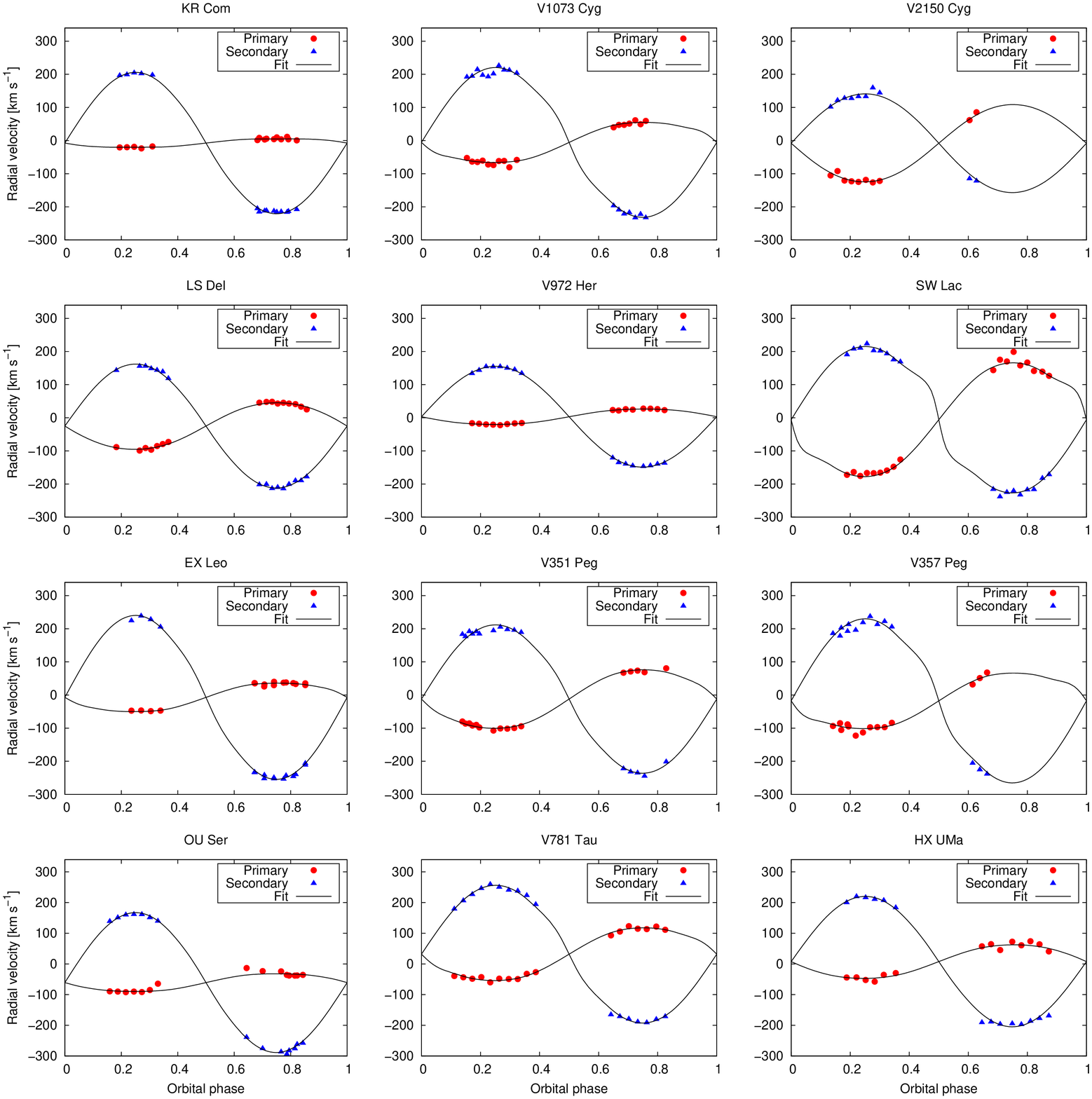}
   \caption{Radial velocity curves of every object with the fitted \texttt{PHOEBE} models. We note that the formal errors of the single velocity points are smaller than their symbols.}
   \label{rv_all}
   \end{figure*}

\section{Sample of observed and synthetic spectra for the observed objects}

   \begin{figure*}
   \centering
   \includegraphics[width=17cm]{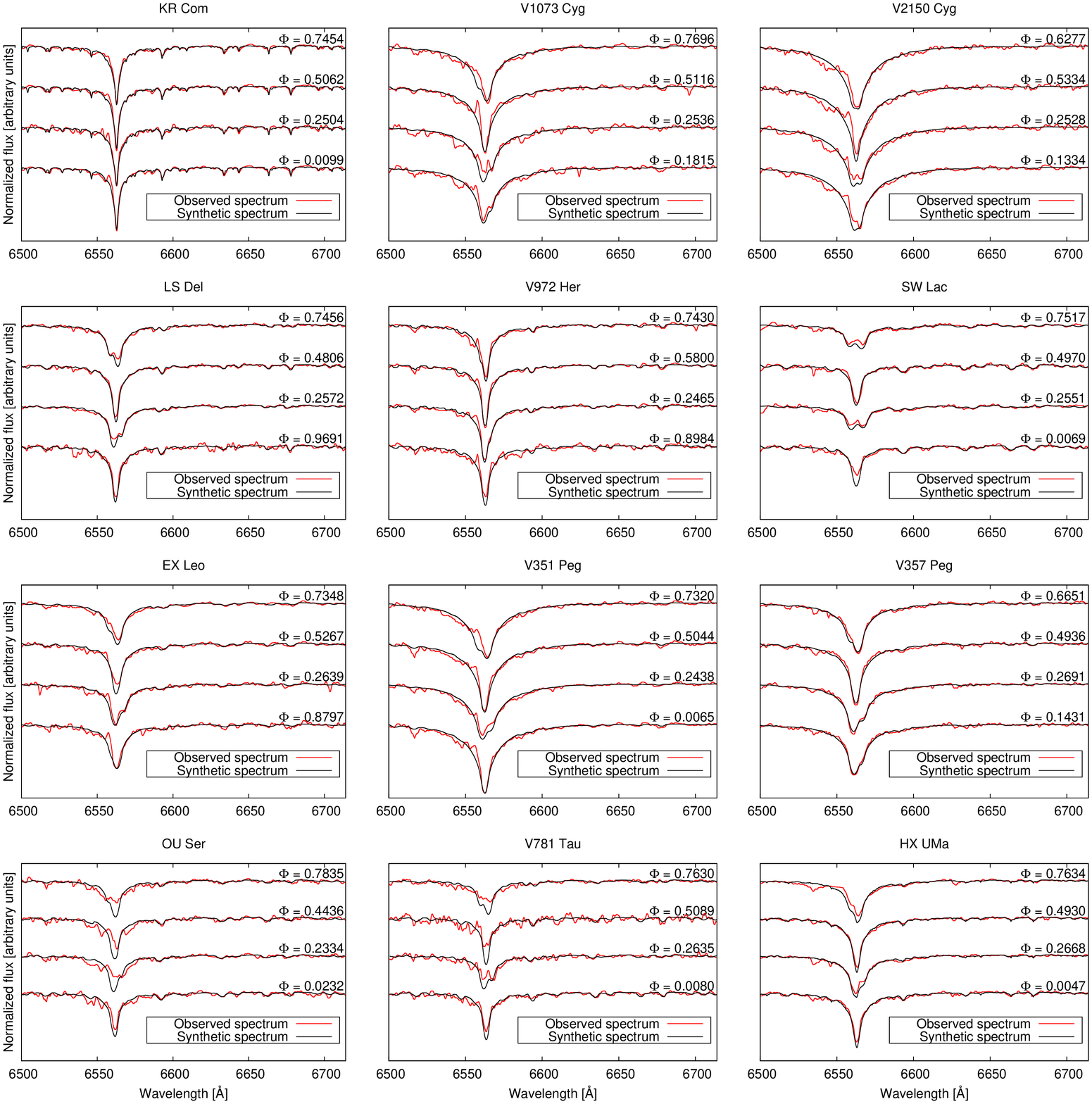}
   \caption{A sample of observed (red line) and synthesized (black line) spectra of every object at different orbital phases.}
   \label{models_all}
   \end{figure*}

\section{Derivation of the Rossby numbers}

The Rossby numbers were calculated as the orbital period ($P$) divided by the convective turnover time ($\tau_{c}$):
\begin{equation}
    R_{0}=\frac{P}{\tau_{c}}.
\end{equation}
The convective turnover times were derived by the formula of \citet{noyes84}:

\begin{equation}
    \log\tau_{c}=    \begin{cases}
      1.362-0.166x+0.025x^{2}-5.323x^{3}, & x>0 \\
      1.362-0.14x, & x<0,
    \end{cases}
\end{equation}
where $x=1-(B-V).$
\end{appendix}

\end{document}